\documentstyle[11pt,newpasp,twoside,epsf]{article}
\def\edcomment#1{\iffalse\marginpar{\raggedright\sl#1\/}\else\relax\fi}
\reversemarginpar

\begin{document}
\title{Theoretical and Observational Agreement on Mass
Dependence of Cluster Life Times}

\author{Mark Gieles$^1$, Holger Baumgardt$^2$, Nate Bastian$^{1,3}$, Henny Lamers$^{1,4}$}
\affil{$^1$ Astronomical Institute, Utrecht University, Princetonplein 5,
3584 CC, Utrecht, The Netherlands\\
$^2$ Institute of Advanced Physical and Chemical Research  RIKEN,
2-1 Hirosawa, Wako-shi, Saitama 351-019, Japan
\\
$^3$ European Southern Observatory, Karl-Schwarzschild-Strasse 2
D-85748 Garching b. M\"unchen, Germany\\
$^4$ SRON Laboratory for Space Research, Utrecht, The Netherlands}

\begin{abstract}
Observations and $N$-body simulations both support a simple relation for the
disruption time of a cluster as a function of its mass of the form:
$t_{\rm dis} = t_{4}\times(M/10^{4}\,M_{\odot})^{\gamma}$.  The
scaling factor $t_{4}$ seems to depend strongly on the
environment. Predictions and observations show that $\gamma \simeq
0.64 \pm 0.06$. Assuming that $t_{\rm dis} \propto M^{0.64}$ is caused
by evaporation and shocking implies a relation between the radius and
the mass of a cluster of the form: $r_{\rm h} \propto M^{0.07}$, which
has been observed in a few galaxies. The suggested relation for the
disruption time implies that the lower mass end of the cluster initial
mass function will be disrupted faster than the higher mass end, which
is needed to evolve a young power law shaped mass function into the
log-normal mass function of old (globular) clusters.

\end{abstract}

\section{Introduction}
	This study combines the results of theoretical and
     observational studies to the life time of stellar clusters. It is
     of vital importance to understand how disruption processes
     influence a cluster population in order to get better
     understanding of the observed luminosity and mass distribution. A
     key question in studying cluster mass functions is whether the
     power law mass distribution of young clusters, as observed in
     interacting/merging and starburst galaxies, will evolve into a
     log-normal mass distribution as observed for old globular cluster
     populations. If so, a preferential depletion of low mass clusters
     is necessary. The two most important dynamical processes
     influencing the cluster mass function are 1) evaporation, which
     is due to internal relaxation of the cluster and 2) tidal
     shocking. The first is a result of two body interactions within the
     cluster which drive the low mass stars through interaction with
     higher mass stars to the edge of the cluster until they reach a
     velocity greater than the escape velocity and leave the
     cluster. Tidal shocks can be caused by passing giant molecular
     clouds (GMCs) or when the cluster travels through a disk, bulge
     or spiral arm. The stars in the cluster gain kinetic energy which
     causes the cluster to expand and become less gravitationally
     bound.

     \section{Analytical expressions for the disruption time}
		
%
%

 The time scales involved for relaxation and shocks depend on the mass
 and the half-mass radius of the cluster: $t_{\rm rh} \propto
 M^{1/2}\,r_{\rm h}^{3/2}$ and $t_{\rm sh} \propto M\,r_{\rm
 h}^{-3}$. Boutloukos \& Lamers (2003) (BL03) and Zepf et al. (1999)
 have proposed that if a general dependence between the disruption
 time and the mass exists of the form $t_{\rm dis} \propto
 M^{\gamma}$, that there should be a relation between the mass and the
 radius of the cluster: $r_{\rm h} \propto M^{0.07}$ in order to
 reduce both time scales to the simple expression $t_{\rm rh} \propto t_{\rm sh}
 \propto M^{\sim2/3}$. This shallow relation between radius and mass
 has been observed for clusters in NGC3256 (Zepf et al. 1999) and
 several spiral galaxies (Larsen, these proceedings)

	\section{Empirical determination of disruption time}
	BL03 propose a scaling law for the disruption time of a star cluster:
	
	\begin{equation}
	t_{\rm dis} = t_{4}\times(M/10^{4}\,M_{\odot})^{\gamma}
	\end{equation}
	with $t_{4}$ the typical disruption time of a 10$^{4}
	M_{\odot}$ cluster and $\gamma$ a dimensionless index. $t_{4}$
	and $\gamma$ can be derived from the age and mass
	distributions of a cluster sample. BL03 did this for different
	galaxies and found that $t_4$ is varying per galaxy but that
	$\gamma$ is remarkably constant with $\gamma \simeq
	0.62\pm0.06$. Disruption according to Eq.~1 is needed to
	explain the observed mass and age distribution of clusters in
	M51 (see also Gieles et al., these proceedings)

	\section{Disruption of clusters with $N$-body simulations}
	Baumgardt \& Makino (2003) have studied disruption of clusters
	of different mass and in different orbits in an external tidal
	field by means of $N$-body simulations. A clear power-law
	relation between the disruption time and the initial mass is
	found. The mean predicted slope, corresponding to the $\gamma$
	value, is $0.66 \pm 0.04$ which agrees very well with the
	empirical value. The scaling value $t_{4}$ found in these
	simulations depends strongly on the distance to the galactic
	center and ranges between 2 Gyr and 10 Gyr for clusters in a
	spherical symmetric potential. This is much larger then the values found
	by BL03 ($\pm$1 Gyr for the solar neighborhood). This can be
	explained by the different definitions of the empirical
	disruption time and the calculated life time and by the fact
	that the simulations did not include a disk, spiral arms and
	GMCs which all decrease the life time of the cluster in the
	disk of our galaxy.


\begin{references}
\reference {Baumgardt}, H.~\& {Makino}, J.\ 2003, MNRAS, 340, 227 
\reference {Boutloukos}, S.~G.~\& {Lamers}, H.~J.~G.~L.~M., 2003, MNRAS 338, 717 
\reference {Zepf}, S.~E., {Ashman}, K.~M., {English}, J., {Freeman}, K.~C., \& {Sharples}, R.~M.\ 1999, AJ, 118, 752 

\end{references}
\end{document}